\documentclass{article}
\usepackage[utf8]{inputenc}
\usepackage{authblk}
\usepackage{hyperref}
\usepackage{bm}

\title {Maxwell's demon and \emph{impossibility statements}: Einstein on perpetuum mobile of the second kind}
\author{Galina Weinstein}
\affil{\normalsize Reichman University, The Efi Arazi School of Computer Science, Herzliya; University of Haifa, The Department of Philosophy, Haifa, Israel.} 

\begin{document}

\maketitle

\maketitle

\begin{abstract}
This paper discusses Maxwell’s demon thought experiment. In recent years, there has been extensive research on Maxwell’s demon. I first provide a detailed overview of important historical milestones of the Maxwell's demon thought experiment. Einstein would often write about both relativity and thermodynamics. In this paper I first argue that: 1. Einstein spoke of a perpetuum mobile of the second kind (Maxwell's demon) in the context of the discovery of his relativity theory, the famous chasing a light beam thought experiment, and the contraction of lengths. 2. He would often compare relativity with thermodynamics. 3. When doing so, he would say that perpetual motion machines cannot exist. In the same category, no signal transmission faster than light is possible. 4. Einstein's approach in speaking of the principles of thermodynamics and relativity suggests that he was occupied with \emph{impossibility statements}. After discussing the essential aspects of Maxwell's demon and the different ways of exorcising the demon, I examine the role of \emph{impossibility statements} in Maxwell's demon thought experiment.
\end{abstract}

\section{Introduction}

This paper discusses the Maxwell’s demon thought experiment. Leo Szilard asks what happens when “an intelligent being intervenes in a thermodynamic system”? He answers that a perpetual motion machine akin to Maxwell's demon is possible if "according to the general methods of physics – we view the experimenting man as a sort of \emph{deus ex machina}” (\cite{Szilard2}). Ian Hacking resists calling Maxwell's demon a thought experiment. He argues that the demon "is part of a fantasy" because "it is hard to see what is experimental about the demon. Perhaps it is only a rhetorical device to reinforce Maxwell's statistical analysis. The demon does not, for me, prove even the possibility of anything. Perhaps it shows an aspect of a scene in statistical mechanics. […] The problem with Maxwell's demon as an experiment is that you can't conduct it at all, no matter how much fantasy and idealization you allow yourself (\cite{Hacking}). Without entering into the debate about thought experiments and experiments, in this paper, I use the phrase Maxwell's demon \emph{thought experiment}. 

Maxwell's demon has caught the imagination of many people. 
Consider an isolated closed system. Maxwell’s demon sits at the partition between two chambers of gas and opens and closes a door between the two chambers. When a fast-moving molecule approaches from one side Maxwell’s doorkeeper opens the trap door, allowing only fast-moving molecules through into the first chamber and slow-moving ones into the second. After doing this many times, the first chamber becomes hotter while the second cools down. Since the average kinetic energy of gas particles is proportional to the absolute temperature of the gas, the total entropy of the gas in the container is decreased, in apparent violation of the second law of thermodynamics. The demon is measuring molecules (separating molecules) without expending any energy (i.e. without performing work), and mechanical work is extracted from the temperature difference between the two chambers. This is a perpetual motion machine of the second type. Maxwell, however, said that he invented his thought experiment to show that the Second law is a statistical law and not a dynamical certainty (\cite{Klein}). 

The Maxwell’s demon problem has been the subject of many classic studies in history and philosophy of science. One of the most significant current discussions in the research of Maxwell’s demon thought experiment is whether the second law of thermodynamics universally valid. Many attempts have been made to exorcise the demon (i.e., show why the demon does not violate the Second Law), and philosophers have been occupied with reflection on the nature of the demon (see \cite{Bub}, \cite{Feyerabend}, \cite{Maroney}, \cite{Earman1} and \cite{Earman2}, \cite{Popper1}). According to John Earman and John Norton, the Bennett-Landauer approach (see section \ref{4}) cannot provide a reliable exorcism of the Demon (\cite{Earman2}). Meir Hemmo and Orly Shenker, however, most conspicuously argue that the Maxwell’s demon thought experiment shows that the Second Law is not universally valid in a strong sense and claim that the demon is compatible with all known principles of classical (and quantum) mechanics (\cite{Shenker1}; \cite{Shenker2}; \cite{Shenker3}; and \cite{Pitowsky}). In this paper I do not intend to defend any of the views presented in the above-mentioned papers; nor do I have any quarrel with these ideas.

In the first four sections of the paper (\ref{1}, \ref{2}, \ref{3}, \ref{4}), I provide an overview of important historical milestones related to Maxwell's demon. 
Considerable efforts have been made to invent mechanisms that show why the Maxwell’s demon thought experiment does not violate the Second Law, and new versions of the original thought experiment have been invented, to name just three known ones: Szilard’s engine (\cite{Szilard1}), Maryan von Smoluchowski’s argument that the demon could be conceived as a trapdoor; and Feynman’s gadget (ratchet and pawl) which generates work from a heat reservoir with everything at the same temperature (\cite{Smoluchowski}; \cite{Feynman}). 

The concept of “seeing” and the requirement of getting information about the position and velocity of the molecules is central to the Maxwell’s demon thought experiment. It was shown that this is problematic according to the Heisenberg uncertainty principle (\cite{Slater}). Because the demon is a measuring device, and its actions depend on the results of the measurement (\cite{Zaslavsky}), Léon Brillouin suggested the use of light signals to obtain information about the molecules (\cite{Brillouin1}, \cite{Brillouin2}). I will end my overview with: Landauer's erasure principle (\cite{Landauer}), Bennett's accounting of the demon's cycle of operation showing that the Maxwell's demon can make its measurement without net energy dissipation, and quantum error correction 
(\cite{Bennet1}, \cite{Bennet2}, \cite{Bennet3}; \cite{Bennet2}, \cite{Bennet5}; \cite{Nielsen}). 

In section \ref{5}, I discuss Maxwell's demon, thermodynamics and relativity. On several occasions, Einstein expressed his opinion on thermodynamics and perpetual motion. I argue:

1. Einstein spoke of a perpetuum mobile of the second kind (Maxwell's demon) in the context of the discovery of his relativity theory, the famous chasing a light beam thought experiment, and the contraction of lengths (\cite{Einstein2}; \cite{Einstein6}; \cite{CPAE3}). 

2. He would often compare relativity with thermodynamics. 

3. When doing so, he would say that perpetual motion machines cannot exist. In the same category, no signal transmission faster than light is possible. 

4. Einstein's approach in speaking of the principles of thermodynamics and relativity suggests that he was concerned with \emph{impossibility statements}. 
I use the phraseology of Scott Aaronson: "Many of the deepest principles in physics are impossibility statements: for example, no superluminal signalling and no perpetual motion machines" (\cite{Aaronson}). 

In section \ref{5}, I then examine the role of \emph{impossibility statements} in Maxwell's demon thought experiment.

\section{Maxwell invents the demon} \label{1}

When Peter Guthrie Tait was preparing his book, \emph{Sketch of Thermodynamics} (published in 1868), for the press he wrote to Maxwell asking him for some corrections. 
Maxwell replied in a letter dated December 11, 1867. In this Letter, it is probably the first time that Maxwell mentioned “Maxwell's Demon” - as William Thomson (Lord Kelvin) called him - “who operating on the individual molecules of a gas could render nugatory the second law of thermodynamics”. Maxwell told Tait that he wanted "To pick a hole" in the second law of thermodynamics.

Maxwell wrote to Tait: "Let $A$ and $B$ be two vessels divided by a diaphragm and let them contain elastic molecules in a state of agitation which strike each other and the sides" of the vessel. "Conceive a finite being who knows the paths and velocities of all the molecules by simple inspection but who can do no work except open and close a hole in the diaphragm by means of a slide without mass". Let the being "first observe the molecules in $A$ and when he sees one coming" whose $v^2$ [$v$ = velocity] is $<$ than the mean-square velocity of the molecules in $B$ "let him open the hole and let it go into $B$. Next let him watch for a molecule of $B$", whose $v^2$ is greater than the mean-square velocity of the molecules in $A$, "and when it comes to the hole let him draw the slide and let it go into $A$, keeping the slide shut for all other molecules” (\cite{Knott}).    

In 1870, in a letter to John Willian Strutt (Lord Rayleigh), Maxwell again wrote about the being: 
Consider the kinetic theory of gases. The different molecules in a gas at uniform temperature are moving with very different velocities. “Put such a gas into a vessel with two compartments and make a small hole in the wall about the right size to let one molecule through. Provide a lid or stopper for this hole and appoint a doorkeeper, very intelligent and exceedingly quick, with microscopic eyes but still an essentially finite being. Whenever he sees a molecule of great velocity coming against the door from $A$ into $B$ he is to let it through, but if the molecule happens to be going slow he is to keep the door shut. He is also to let slow molecules pass from $B$ to $A$ but not fast ones". Maxwell stresses that "Of course he must be quick for the molecules are continually changing both their courses and their velocities. In this way the temperature of $B$ may be raised and that of $A$ lowered without any expenditure of work, but only by the intelligent action of a mere guiding agent” (\cite{Harman}).    
In other words, $B$ becomes hot and $A$ cools, with a negligible expenditure of work; and the demon cools the gas indefinitely. 

In 1871 the following now-classic text by Maxwell appeared in print (\cite{Maxwell1}):

\begin{quote}
“But if we conceive a being whose faculties are so sharpened that he can follow every molecule in its course, such a being, whose attributes are still as essentially finite as our own, would be able to do what is at present impossible to us. For we have seen that the molecules in a vessel full of air at uniform temperature are moving with velocities by no means uniform, though the mean velocity of any great number of them, arbitrarily selected, is almost exactly uniform. Now let us suppose that such a vessel is divided into two portions, $A$ and $B$, by a division in which there is a small hole, and that a being, who can see the individual molecules, opens and closes this hole, so as to allow only the swifter molecules to pass from $A$ to $B$, and only the slower ones to pass from $B$ to $A$. He will thus, without expenditure of work, raise the temperature of $B$ and lower that of $A$, in contradiction to the second law of thermodynamics”.     
\end{quote}

As said above, in 1874 Thomson famously called Maxwell’s being a demon: “This process of diffusion could be perfectly prevented by an army of Maxwell’s ‘intelligent demons’* stationed at the surface, or interface [...] separating the hot from the cold part of the bar” (\cite{Thomson1}; See \cite{Daub}).

In an undated letter, Maxwell summarized the following characteristics of his demon (\cite{Knott}):

\begin{quote}
"I. Who gave them [the demons] this name? Thomson.

2. What were they by nature? Very small BUT lively beings incapable of
doing work but able to open and shut valves which move without friction or
inertia.

3. What was their chief end? To show that the $2^{nd}$ Law of Thermodynamics has only a statistical certainty".    
\end{quote}

Hence, Maxwell's demon is incapable of doing work and Maxwell's purpose is to show that the second law of thermodynamics is a statistical law. That is, the second law does not have absolute validity; it only has a statistical validity.
In 1877, Maxwell made it clear that his being “could not turn any of the energies of nature to his own account, or to one who could trace the motion of every molecule and seize it at the right moment” (\cite{Maxwell2}).

In 1879 Thomson further described the demon as "a being with no preternatural qualities" who "differs from real living animals only in extreme smallness and agility. He can at pleasure stop, or strike, or push, or pull any single atom of matter, and so moderate its natural course of motion. Endowed ideally with arms and hands and fingers-two hands and ten fingers suffice - he can do as much for atoms as a pianoforte player can do for the keys of the piano - just a little more, he can push or pull each atom in any direction”. The demon can sort the molecules and “separate the gases into different parts of the containing vessel” (\cite{Thomson2}).

\section{Recurrence and Reversibility} \label{2}

The important theoretical developments in the nineteenth century, including thermodynamics, conservation of energy, the kinetic theory of gases, statistical mechanics and electrodynamics, brought physicists to ask: what was the nature of the relationship between these theories and mechanics; asking especially which theory constituted the foundation of physics? This led to a re-examination of the foundations of physics. Several world-pictures were suitable for establishing the foundations of physics: the electromagnetic world-picture, the energetic world-picture, and the mechanistic world-picture. Each worldview had its proponents and advancers. 

In 1893 Henri Poincaré realized that the advocates of the mechanistic worldview have met with several obstacles in their attempts to reconcile mechanics with experience. Because according to the mechanistic world-picture, all phenomena must be reversible, while experience shows that many phenomena are irreversible. It has been suggested that the apparent irreversibility of natural phenomena is due merely to the fact that molecules are too small and too numerous for our macroscopic senses to deal with them. "Although", says Poincaré, "a ‘Maxwell demon’ could do so and would thereby be able to prevent irreversibility”. 

But Poincaré used his recurrence theorem to remove the need for the demon. According to Poincaré's recurrence theorem, the entropy decreases when the initial state eventually recurs (repeats itself): “A theorem, easy to prove, tells us that a bounded world, governed only by the laws of mechanics, will always pass through a state very close to its initial state”. The recurrence theorem takes into consideration that the number of molecules in the world is finite. If this is so, then there is only a finite number of possible arrangements of molecules; and if time is infinite, the same combination of molecules is bound to come up again. At some instant, all the molecules in the world would arrange themselves in the same form they had at some previous time. 

Poincaré believed that to get a perpetual motion machine of the second kind, “it will not be necessary to have the acute vision, the intelligence, and the dexterity of Maxwell's demon; it will suffice to have a little patience”. Thus, Poincaré was willing to accept the possibility of a violation of the Second Law after a very long time has passed (\cite{Poincare2}; see \cite{Poincare1}; \cite{Holton}). 

Ludwig Boltzmann formulated the $H$ theorem (where $H \propto -$ entropy) and showed that $H$ must decrease (the entropy would increase). In other words, the same combination of molecules cannot come up again. Because the system of molecules evolves from non-equilibrium to equilibrium and once the system reaches equilibrium, it will not move from equilibrium. 
Ernest Zermelo used Poincaré's recurrence theorem to argue that this might not be the case, because some time in the future, a system of molecules may return to its initial state or to a state arbitrarily close to it. This is called \emph{(Poincaré's) Zermelo's Wiederkehreinwand}.

Boltzmann raised objections to Poincare's recurrence telling Zermelo that there is, however, a great problem with Poincaré's recurrence theorem: the recurrence time is so enormous more than the age of the universe itself. But even if the time between recurrence and initial state were a little shorter, it is still long enough that all information about the initial state is lost. Furthermore: “Poincare's theorem does not contradict the applicability of probability theory but rather supports it, since it shows that in eons of time there will occur a relatively short period during which the state probability and the entropy of the gas will significantly decrease, and that a more ordered state similar to the initial state will occur. During the enormously long period of time before this happens, any noticeable deviation of the entropy from its maximum value is of course very improbable; however, a momentary increase or decrease of entropy is equally probable". Boltzmann concludes: "since one intentionally starts from a very improbable state. In the case of natural processes this is explained by the assumption that one isolates the system of bodies from the universe which is at that time in a very improbable state as a whole” (\cite{Boltzmann2}). 

Zermelo answered Boltzmann, saying that probability has nothing to do with time (\cite{Zermelo}):

\begin{quote}
"I have therefore not been able to convince myself that Herr Boltzmann's probability arguments, on which 'the clear comprehension of the gas-theoretic theorem' is supposed to rest, are in fact able to dispel the doubts of a mechanical explanation of irreversible processes based on Poincaré's theorem, even if one renounces the strict irreversibility in favour of a merely empirical one. Indeed it is clear \emph{a priori} that the probability concept has nothing to do with time and therefore cannot be used to deduce any conclusions about the \emph{direction} of irreversible processes. On the contrary, any such deduction would be equally valid if one interchanged the initial and final states and considered the \emph{reversed} process running in the opposite direction.    
\end{quote}

In 1869, Johann Josef Loschmidt "stated a theorem that casts doubt on the possibility of a purely mechanical proof of" the second law thermodynamics (\cite{Boltzmann1}). This "theorem" represented an attempt to do exactly the kind of thing which Maxwell had suggested to Tait, namely, "to pick a hole" in the second law of thermodynamics. Loschmidt's idea is sometimes called \emph{Loschmidt’s demon} and also known as \emph{Loschmidt’s reversibility paradox} or \emph{Umkehreinwand}. But unlike Maxwell, Loschmidt argued that the true basis of the second law of thermodynamics must be established on dynamical principles. 

Loschmidt objected to Boltzmann’s proof of the increase of entropy: let us assume that an isolated system evolves from an initial state to a final state of lower $H$. Since the microscopic laws of mechanics are invariant under
time reversal, there must also exist an entropy decreasing evolution for which $H$ increases with time. That evolution, is, of course, set in motion simply by taking the final state of the previous evolution as the new initial state and then reversing all the individual molecular velocities. This time-reversed evolution would seem to violate the second law of thermodynamics. 

Boltzmann set himself to answer Loschmidt’s reversibility objection. What Loschmidt showed was that the second law of thermodynamics cannot be absolute (if the second law is absolute, then $H$ can never increase). Boltzmann replied to Loschmidt on the contradiction between the irreversibility of thermodynamic processes and the reversibility of the underlying mechanical laws. Boltzmann suggested a probabilistic explanation for the empirical irreversibility, embodied in the second law of thermodynamics. 
According to Boltzmann, one cannot prove that whatever may be the initial positions and velocities of the molecules, their distribution must become uniform after a long time; rather one can only prove that infinitely many more initial states will lead to a uniform distribution after a definite long time than to a non-uniform one. Thus, Boltzmann tells Loschmidt that, it is more probable that the gas will be spread uniformly than grouped in one corner of the chamber. Boltzmann goes on to say: Loschmidt's theorem tells us only about initial states which lead to a very non-uniform distribution of states after a certain time; but it does not prove that there are not infinitely many more initial conditions that will lead to a uniform distribution after the same time.
In other words, of all conceivable arrangements of the gas molecules at a given instant, their motion would nearly always be completely random and uniform. Yet, Boltzmann agrees with Loschmidt that, it is certainly probable that at some moment most of the molecules happen, by chance, to be moving in the same direction. Fluctuations from complete uniform motion are bound to occur. But for collections of molecules as large as $10^{23}$, the chance of a fluctuation large enough to be measurable is extremely small, but certainly not zero.

Jos Uffink points out that according to Boltzmann, the initial conditions considered by Loschmidt only have an extraordinarily small probability. Consequently, from Boltzmann's point of view, Loschmidt’s state is practically impossible (\cite{Uffink}).

Loschmidt’s reversibility objection was one of the factors that motivated Boltzmann to support Maxwell's view. Recall that Maxwell invoked his demon "to show that the $2^{nd}$ Law of Thermodynamics has only a statistical certainty". It deals with probabilities, not certainties (\cite{Knott}; \cite{Boltzmann1}; \cite{Boltzmann2}; see \cite{Canales}; \cite{Daub} and \cite{Leff1}; \cite{Bader}; \cite{Holton}).  

Thomson raised similar concerns to those raised by Loschmidt: “In abstract dynamics an instantaneous reversal of the motion of every moving particle of a system causes the system to move backwards, each particle of it along its old path, and at the same speed as before when again in the same position – that is to say, in mathematical language, any solution remains a solution when $t$ is changed to $-t$". Thomson adds that in mechanics, "this simple and perfect reversibility fails" on account of disturbing forces. "If, then, the motion of every particle of matter in the universe were precisely reversed at any instant, the course of nature would be simply reversed for ever after” (\cite{Thomson1}).    
And in 1870, in the letter to Strutt, Maxwell also wrote about reversibility (\cite{Harman}): “Dear Strutt, if this world is a purely dynamical system and if you accurately reverse the motion of every particle of it at the same instant then all things will happen backwards. The possibility of executing this experiment is doubtful but I do not think that it requires such a feat to upset the $2^{nd}$ law of Thermodynamics". All we need is a demon. 

Boltzmann emphasized the close relationship between the second law of thermodynamics and probability theory. But as Zermelo stressed to Boltzmann, probability theory has nothing to do with time. Boltzmann linked between entropy and probability, and wrote in 1877 that, a system evolves to the most probable state, that of equilibrium, and "If we apply this to the second law, then we can identify that quantity, which one usually calls the entropy, with the probability of the state in question" (\cite{Boltzmann3}; \cite{Uffink}).  

In his \emph{Autobiographical Notes}, Einstein wonderfully explained Boltzmann's probabilistic interpretation of the entropy: 

\begin{quote}
“On the basis of the kinetic theory of gases Boltzmann had discovered that, aside from a constant factor, entropy is equivalent to the logarithm of the 'probability' of the state under consideration. Through this insight he recognized the nature of course of events which, in the sense
of thermodynamics, are 'irreversible'. Seen from the molecular-mechanical point of view, however all courses of events are reversible. If one calls a molecular-theoretically defined state a microscopically described one, or, more briefly, micro-state, and a state described in terms of thermodynamics
a macro-state, then an immensely large number ($Z$) of states belong to a macroscopic condition. $Z$ then is a measure of the probability of a chosen macro-state. This idea appears to be of outstanding importance also because of the fact that its usefulness is not limited to microscopic description on the basis of mechanics".    
\end{quote}

The Boltzmann entropy is a logarithmic measure of the number of microstates $W$ corresponding to the macrostate whose entropy is $S$. In 1905 Einstein called this famous relationship the "Boltzmann principle" (\cite{Einstein7}; \cite{Pais}).

Macroscopic thermodynamic properties (temperature, pressure, and volume) are represented by the microstates, which macroscopically all look the same, but are different microscopically, in their position and velocity. 
The number of particles (or degrees of freedom) in a macroscopic system is characterized by Avogadro’s number $N=10^{23}$. Consider a system of $N$ particles or molecules that macroscopically fills out half of the volume of the box. The overwhelming majority of microstates would evolve into what will look, macroscopically, like a gas that is homogeneously distributed over the entire volume of the box. The phase space volume corresponding to this equilibrium macrostate consists almost entirely of phase points whose totality has a volume of order $10^N$. In the Boltzmannian picture, a non-equilibrium microstate occupies an extremely small region of phase space while equilibrium microstates occupy almost the entire phase-space volume. The overwhelming majority of microstates move from non-equilibrium to equilibrium, and so entropy is low in the past and high in the future, and not the other way round.  
The dynamics of a system of $N=10^{23}$ particles are very chaotic and unstable. Even small variations in the initial conditions can lead to considerable differences in the evolution of the system. Such a large number of particles plus the chaotic nature of the dynamics explain why it is almost impossible for a system to be in a microstate which will evolve in isolation in a way contrary to the second law of thermodynamics.  
Real systems are not perfectly isolated. The effect of unavoidable small external influences will greatly destabilize the evolution of the system.

A modern version of Boltzmann’s approach to statistical mechanics is typicality (\cite{Allori};\cite{Goldstein};\cite{Lebowitz1}, \cite{Lebowitz2}, \cite{Frigg} \cite{Lazar}). Typicality is not a probability. Microstates of phase space evolve towards a larger phase space volume, i.e., towards an equilibrium macroscopic state, and this is a typical behavior. 
The dynamics is irrelevant and there is no need to consider the Hamiltonian equations of motion. That is because equilibrium depends on the structure of the phase space, the size of the equilibrium and not on the dynamics. 
Thus, typicality is a claim valid only for the overwhelming majority of initial conditions. All sorts of objections were raised against typicality (see \cite{Hagar}, \cite{Pitowsky2}, \cite{Shenker2}). For instance,  Uffink argues: "States don’t evolve into other states just because there are more of the latter, or because they make up a set of larger measure. The evolution of a system depends only on its initial state and its Hamiltonian. Questions about evolution can only be answered by means of an appeal to dynamics, not by the measure of sets alone. [...] The lesson is, of course, that in order to obtain any satisfactory argument why the system should tend to evolve from non-equilibrium states to the equilibrium state, we should make some assumptions about its dynamics" (\cite{Uffink}). 

\section{The demon makes measurements} \label{3}

How can Maxwell’s demon observe and see the molecules, and watch for a molecule?

In 1941, Percy Bridgman wrote: “Another doubtful feature is the method by which the demon would learn of the approach of the individual molecules. The only method would appear to be by light signals; these must come in quanta and must react with the molecule.  The reaction is uncontrollable, and may be sufficiently large to divert the molecule by so much as to vitiate the manipulations of the trap door” (\cite{Bridgman}, p. 157). 

In 1944 Pierre Demers suggested that “the demon can ascertain the presence of a molecule, its nature, and its state of motion" by using "signals of light”. Thus, Demers proposed a “special process for optically measuring the speed of a molecule” by sending light signals to the molecules (\cite{Demers}). 

Recall that Maxwell wrote in 1871: “a being, who can see the individual molecules” (\cite{Maxwell1}). In other words, to count the molecules the demon must “see” them. 

Brillouin closely followed the work of Demers and assumed the use of photons. The demon “sees” the particles by reflecting photons off them. But these reflected photons must be distinguished from the background noise of the blackbody radiation (\cite{Leff2}; \cite{Finfgeld}). 

But Brillouin was probably also inspired by Bridgman. In 1949 Brillouin wrote: "In a discussion at Harvard (1946), P. W. Bridgman stated a fundamental difficulty regarding the possibility of applying the laws of thermodynamics to any system containing living organisms. How can we compute or even evaluate the entropy of a living being? [...] The entropy content of a living organism is a completely meaningless notion. In the discussion of all experiments involving living organisms, biologists always avoid the difficulty by assuming that the entropy of the living objects remains practically constant during the operation. This assumption is supported by experimental results, but it is a bold hypothesis and impossible to verify. [...] There are other remarkable properties characterizing the ways of living creatures. For instance, let us recall the paradox of Maxwell's demon, that submicroscopical being, standing by a trapdoor and opening it only for fast molecules, thereby selecting molecules with highest energy and temperature. Such an action is unthinkable, on the sub microscopical scale, as contrary to the second principle$.^{11}$" (\cite{Brillouin1}).

The first time that Brillouin suggested that the demon "sees" the molecules with a torchlight was in the comment $11$ (after the words "as contrary to the second principle", in the above quotation) (\cite{Brillouin1}): 

\begin{quote}
“In order to choose the fast molecules, the demon should be able to see them; but he is in an enclosure in equilibrium at constant temperature, where the radiation must be that of the black body, and it is impossible to see anything in the interior of a black body. The demon simply does not see the particles, unless we equip him with a torchlight, and a torchlight is obviously a source of radiation not at equilibrium. It pours negative entropy into the system. Under these circumstances, the demon can certainly extract some fraction of this negative entropy by using his gate at convenient times. Once we equip the demon with a torchlight, we may also add some photoelectric cells and design an automatic system to do the work”.    
\end{quote}

Two years later, Brillouin asked: “Is it actually possible for the demon to see the individual atoms?” As evident from the above quotation, Brillouin interpreted Maxwell’s being as follows: to properly select the molecules, the demon must be capable of seeing them. That is, he must detect and measure their position and velocity. But the demon is in an isolated system at thermal equilibrium, where the radiation is blackbody radiation; he cannot see anything in the interior of a blackbody, let alone observe the molecules. Hence, the demon cannot operate the trap door and he is unable to violate the Second Law. 

Brillouin exorcises the demon by equipping him with an “electric torch so that he can see the molecules”. Brillouin’s “torch is a source of radiation not in equilibrium. It pours negative entropy into the system. From this negative entropy the demon obtains ‘informations’. With these informations he may operate the trap door and rebuild negative entropy, hence, completing a cycle: Negentropy $\rightarrow$ information $\rightarrow$ negentropy. We coined the abbreviation ‘negentropy’ to characterize entropy with the opposite sign. […] Entropy must always increase, and negentropy always decreases” (\cite{Brillouin2}). Brillouin shows that there is an intimate link between entropy and information.

According to quantum physics, the demon will have to expend a minimum of one photon to see the molecules, and the molecule to be observed must also scatter at least one photon. There is a caveat though, if the demon wishes to see anything at all, the photon must be of short-wavelength and thus more energetic than the background thermal radiation (because the energy of a photon depends on its wavelength). The demon can now decrease the entropy of the system of gas and receive a certain amount of information about the molecules, but the entropy of the system increases, and the Second Law is not violated. 
Brillouin concludes his discussion by noting: “As [Dennis] Gabor states it: ‘We cannot get anything for nothing, not even an observation’” (\cite{Brillouin2}; \cite{Brillouin3}; \cite{Carnap}).

Szilard’s engine (described in the next section \ref{4}) inspired Gabor to devise a machine, a perpetuum mobile of the second kind, and then use a light beam to save the Second Law: “the molecule drifts into the light stream. A part of the light will be scattered and collected by one or the other or both photosensitive elements” (\cite{Gabor}).  

Brillouin mentions a problem with detecting and measuring simultaneously the position and velocity of molecules with infinite accuracy. In 1939, John Clarke Slater warned against the possibility of imagining demons like the one proposed by Maxwell: "Is it possible, we may well ask, to
imagine demons of any desired degree of refinement? If it is, we can make any arbitrary process reversible, keep its entropy from increasing, and the second law of thermodynamics will cease to have any significance. The answer to this question given by the quantum theory is No”. The reason is that according to the Heisenberg's principle of uncertainty, the two quantities, position and velocity of each molecule, cannot be measured simultaneously with infinite accuracy (\cite{Slater}; \cite{Brillouin3}).  

William Poundstone describes Brillouin’s Maxwell’s demon with a lamp in a picturesque way: if the demon's eyes are sensitive to infrared light, he will find the interior of his sealed chamber bathed in a warm glow. Short wavelength photons (in the infrared range) are energetic photons. Infrared photons will bounce off the molecules and into the demon's eyes. He will see the molecules. The demon's chamber can hold two molecules or a quintillion. Blackody photons' "message" is already as garbled as it can be. All the demon will ever see is a random swarm of photons. To see anything, the demon must provide his own light. It must be a coherent beam of light whose initial direction is known so that a change in direction (by reflection off a molecule) is detectable. The filament of the demon's flashlight must be at a higher temperature than the chamber so that reflected photons will not be confused with the blackbody radiation from the chamber itself (\cite{Poundstone}). 

In 1975, Aleksandr Lerner agreed with Brillouin solution: “In order that the ‘demon’ should be able to estimate the movement of the molecules he must at least see them, and for this purpose it is necessary to provide light. However, the source of light is a system which is not in equilibrium and cannot function without expenditure of energy. It has been found that the quantity of energy necessary to obtain the required information will more than absorb the gain from utilizing it, and therefore there is no infringement of the second law of thermodynamics” (\cite{Lerner}).

On the other hand, Olivier Costa de Beauregard raised an objection to Brillouin's solution: “Brillouin [9] claims he has ‘exorcized’ the demon, by showing that, first, to see the swarm of molecules at thermal equilibrium with radiation he must dispose of a flash-light, and that his door cannot be oscillating at this reigning temperature. These are incisive remarks, but they do not solve the problem, because justifying irreversibility by invoking irreversibility is circular” (\cite{Costa de Beauregard}). 
It was indeed argued against Brillouin’s analysis that, thermodynamic entropy is conditioned by the validity of the Second Law. Hence, Brillouin’s solution to the Maxwell's demon problem assumes the validity of the law it attempts to rescue (\cite{Capek}). 

In 1981 Jack Denur argued that Brillouin’s argument is incomplete. The chamber’s atoms are in constant thermal motion as they emit and absorb radiation. From his stationary position at the door, the demon observes Doppler, blue shifted radiation from atoms moving toward him. Thus, by monitoring the subtle blue shifting of the blackbody radiation, he can anticipate the arrival of high-speed atoms and select them accordingly. This is Denur’s Doppler Maxwell’s demon, a Maxwell demon that utilizes Doppler-shifted radiation from moving molecules as a means of selecting them. “Our Doppler demon has appropriate sensors, properly positioned and aligned, for detecting the frequency, wavelength, power flux per unit area, and/or temperature of the radiation which it receives”, says Denur (\cite{Denur}; \cite{Capek}).

Finally, Smoluchowski replaced the demon with a simple mechanical device, a miniature gate that is designed to swing in only one direction, and hinge has a spring that tends to push the gate closed. The spring keeps the gate shut. Now consider a chamber divided by a partition into two equal parts, containing an equal number of gas molecules in each part. The gate will be opened every time a molecule approaches the gate from the right, but closed if it approaches from the left.
Eventually all the molecules are trapped on the left-hand side, the gas is compressed, and the entropy is reduced. If there were no compensating increase of entropy of the mechanism, then the second law of thermodynamics would be violated. 
Alas, although the one-way gate seems to work just like Maxwell’s demon, when the spring pushes the opened door shut, it must simply bounce open again. Very soon, after being struck by a few molecules, the gate will swing open and shut in a quite random way, participating in the thermal motion of the system. Now it is no longer a one-way gate at all. 

Indeed, Brillouin stresses that Smoluchowski had noticed that the possible effect of random unpredictable Brownian motions of the trapdoor, which would result in random opening or closing of the gate could seriously perturb the operation of the perpetual mobile machine of the second kind. In fact, says Brillouin, this is of special importance if the system is operated by an automatic device such as a spring, rather than by a demon (\cite{Brillouin3}; \cite{Bennet4}). Feynman proposed a version of Smoluchowski's thought experiment in his \emph{Feynman’s Lectures on Physics}: the \emph{ratchet and pawl} thought experiment (\cite{Feynman}). 

\section{The demon records, stores and erases information} \label{4}

In 1961, Rolf Landauer argued that information obeys the laws of thermodynamics. Information is stored in a memory and quantified by the information entropy $kT \ln(2)$, which corresponds to the smallest
amount of information, a bit. The larger the entropy, the bigger the information content. Landauer pointed out that \emph{according to Brillouin}, the measurement process requires energy dissipation of bits from the system to the environment: to get one bit of information about a molecule, we have to dissipate an energy of $kT \ln(2)$, where $k$ is Boltzmann’s constant and $T$ is the temperature. Landauer suggested instead that what requires energy dissipation of bits is rather the operation \emph{erasure}. Memory erasure is an irreversible process: an operation that throws away information about the previous logical state of the device, and necessarily generates in the environment an amount of entropy equal to the information thrown away. Landauer called such an operation, \emph{logically irreversible} and said that logical irreversibility, in turn implies physical irreversibility, and the latter is accompanied by dissipative effects: to erase logical information, one must lower the entropy of the device and the entropy of the environment increases. Consequently, heat is dissipated to the environment: erasing one bit of information dissipates an energy of $kT \ln(2)$, and erasure is thus a thermodynamic irreversible operation (\cite{Landauer}; \cite{Bennet5}).

Charles Bennett has invoked the robot demon: the robot demon stands at the trapdoor, programmed to monitor the molecules with video cameras and operates the door as Maxwell’s demon would. The robot could observe the molecules, run its program, and operate the door, all without using any net work. But as it operates, the robot necessarily acquires a memory record of its observations. With billions or trillions of molecules, this record might occupy a lot of storage space. So, according to Landauer's principle, the robot demon would have to erase information and the erasure of information requires work. The robot demon must convert a very small amount of energy into heat (which is expelled into the environment) for each bit of information it erases (\cite{Bennet4}).  

In the 1970s, Bennett began thinking about the thermodynamics of
computation and found that erasure may be thermodynamically reversible. In cases it is reversible, data is erased, and the entropy increase of
the environment is compensated by an entropy decrease of the data, so the operation as a whole is thermodynamically reversible. Bennett argued that "This is the case in resetting Maxwell's demon". 

In 1982, Bennett gave accounting of the demon's cycle of operation and showed that the "measurement is reversible and does not increase the entropy of the universe". He showed that Maxwell's demon can make its measurement without net energy dissipation. 

Bennett considered Szilard's version of the Maxwell's demon thought experiment which uses a gas consisting of a single molecule moving freely in a cylinder: 

1. The demon first inserts a partition trapping the molecule on one side or the other (left or right) of a cylinder. 

2. Next the demon performs a measurement (records and stores information) to learn which side the molecule is on, left or right. 

3. The demon then extracts work by removing the partition and inserting a piston, allowing the molecule to push the piston and expand isothermally to fill the whole container again. 

4. Finally the demon clears its mind in preparation for the next measurement. 

We reset the demon's mind to a blank state in stage 4, which reduces the entropy of the demon from one bit to zero [reduces the entropy of the demon by $kT \ln(2)$]. This step is logically irreversible. By Landauer's principle, the entropy of the environment must increase by one bit. 

This increase cancels the decrease brought about during the previous stage 3, in which the entropy of the environment is reduced by one bit while the entropy of the demon is increased by a bit [by $kT \ln(2)$)] (because the demon records and stores information, stage 2), bringing the cycle to a close with no net entropy change of demon, molecule, or environment (\cite{Bennet2}; \cite{Bennet5}).

Karl Popper has raised objections to Szilard’s thought experiment. First, he argues that he believes "that the second law is actually refuted by Brownian movement". Second, "the assumption of a gas represented by one molecule, $M$, is not only an idealization [...] but amounts to the assumption that the gas is, objectively, constantly in a state of minimum entropy". That is because "the whole $M$ is on one side of the piston only; and so it will push the piston. But assume we have in fact \emph{two} molecules in the gas; then these may be on different sides, and the piston may not be pushed by them. This shows that \emph{the use of one molecule $M$ only} plays an essential role in my answer to Szilard (just as it did in Szilard’s argument) and it also shows that \emph{if} we could have a gas consisting of one powerful molecule $M$, it would indeed violate the second law. But this is not surprising since the second law describes an essentially statistical effect" (\cite{Popper2}). And recall (see section \ref{1}) that this is exactly the reason why Maxwell invented his demon in the first place, namely, to show that the Second Law is a statistical law.

An interesting interpretation of Maxwell’s demon that follows the above Landauer-Bennett approach is the following: 
from a thermodynamic point of view, quantum error-correction may be considered as a type of refrigeration process, capable of keeping a quantum system at a constant entropy, despite the influence of noisy processes in the environment; these tend to change the entropy of the system. Quantum error-correction appears to allow a reduction in entropy of the quantum system in apparent violation of the second law of thermodynamics. 
In other words, quantum error-correction is essentially a special type of Maxwell’s demon. 

We can imagine Maxwell’s demon performing error syndrome measurements (measurements regarding possible errors) on the quantum system. After getting information about the errors, the demon corrects the errors (noise) according to the result of the syndrome measurements. Just as in the analysis of the classical Maxwell’s demon, the storage of the syndrome in the demon’s memory carries with it a thermodynamic cost, in accordance with Landauer’s principle. Because any memory is finite, the demon must eventually begin erasing information from its memory, to have space for new measurement results. Recall that the Landauer erasure principle states that erasing one bit of information from the memory increases the total entropy of the environment by at least one bit. 
More specifically, this can be represented as the following cycle: 

1) The quantum system, starting in a state $\rho$, is subjected to a noisy quantum evolution (a noisy environment) that takes it to a state $\rho^{'}$. 

2) A demon performs a syndrome measurement on the state $\rho^{'}$ obtaining result $m$. The record of this result is kept in the demon’s memory. 

3) The demon applies a recovery operation that creates a final system state equal to the initial state $\rho$. This signifies there might be a decrease in entropy of the quantum system.

4) But the demon then resets its memory by erasing bits, i.e., it discards information and erases the storage of the measurement result, causing a net increase in the entropy of the environment by Landauer’s principle, and the quantum-error correction cycle is restarted.

The process is cyclic, and is a reversible process: the demon returns to its initial state. But it does not recover a particular initial state of the system (see \cite{Bennet1}, \cite{Bennet3}, and \cite{Bennet4}).  

After stage 3, the demon must erase information from its memory. 
The error-correction stages 2) and 3) come at the expense of entropy production in the environment when the measurement record is erased, and the dissipation of bits of entropy into the environment is at least as large as the entropy reduction in the quantum system being corrected (\cite{Nielsen}, \cite{Nielsen2}). 

Vlatko Vedral proposes a somewhat different version of quantum error correction considered as Maxwell's demon akin to the Szilard–Bennett Maxwell's demon gedankenexperiment (\cite{Vedral}, \cite{Vedral2}).

\section{The demon and impossibility statements} \label{5}

Einstein read the works of Boltzmann. He wrote to Mileva Marić: "Boltzmann is absolutely magnificent" (\cite{CPAE1}). In "Lecture Notes for a Course of Kinetic Theory of Heat Delivered at the University of Zurich in the Summer Semester of 1910", Einstein writes: “Obviously, there can be statistical laws for the motion only in the case where the point later on returns arbitrarily close to a point it had already occupied once before” (\cite{CPAE4}). It is highly likely that in writing the above sentence Einstein was inspired by Boltzmann’s response to  \emph{Loschmidt's Umkehreinwand} and \emph{Zermelo's Wiederkehreinwand} (see section \ref{2}). 

Much later in 1949, Einstein explained: “the sending of a signal is, in the sense of thermodynamics, an irreversible process, a process which is connected with the growth of entropy (whereas, according to our present knowledge, all elementary processes are reversible)” (\cite{Einstein1}). For Einstein, this means that signals are seen as related to thermodynamics. One cannot move backwards in time, because no signal transmission faster than light is possible. Likewise, entropy of a system cannot decrease. 

Einstein would often relate relativity to thermodynamics (see \cite{Giovanelli}). He made a distinction between theories of principle, such as thermodynamics, and constructive theories, such as statistical mechanics. He characterized the special theory of relativity as a theory of principle and considered it to be basically complete when the two underlying principles of the theory (the principle of relativity and that of the constancy of velocity of light) were established. All later work would involve the development of constructive theories compatible with these basic principles (\cite{Weinstein}). In 1908, Einstein wrote to Arnold Sommerfeld that it seems to him that a physical theory can be satisfactory only when it is built from elementary foundations. In other words, a physical theory is satisfactory only if it is a principle theory. Einstein went on to say: “The theory of relativity is not more conclusively and satisfactory, than, for example, classical thermodynamics was before Boltzmann had interpreted entropy as probability” (Einstein to Sommerfeld, January 14, 1908, \cite{CPAE5}, Doc. 73). Put differently, classical thermodynamics is a principle theory and Boltzmann’s explanation, i.e., interpretation of entropy as probability constitutes a constructive effort. 

In his \emph{Autobiographical Notes}, Einstein discussed relativity in the context of thermodynamics. In his \emph{Notes}, Einstein recalls that before 1905, he was looking for a general principle for the electrodynamics of moving bodies, a principle of the kind one finds in thermodynamics, the Second Law (\cite{Einstein2}; \cite{Einstein6}): "The example I saw before me was thermodynamics. The general principle was there given in the theorem: the laws of nature are such that it is impossible to construct a \emph{perpetuum mobile} (of the first and second kind). How, then, could such a universal principle be found? After ten years of reflection such a principle resulted from a paradox upon which I had already hit at the age of sixteen: if I pursue a beam of light with the velocity $c$ (velocity of light in a vacuum), I should observe such a beam of light as a spatially oscillatory electromagnetic field at rest. However, there seems to be no such thing, whether on the basis of experience or according to Maxwell's equations". 

In a discussion following a lecture Einstein gave on the theory of relativity, a question was asked: When the second order experiments of Michelson and Morley were conducted, the contraction hypothesis was invoked for the purpose of explaining the apparent constancy of the velocity of light on the moving earth to second order. Hendrik Antoon Lorentz assumed that the change in length is brought about by the influence of the molecular forces. On the other hand, in his famous paper “Space and Time”, Hermann Minkowski wrote that the contraction was dependent on the state of motion and he made it absolutely independent of any physical influence (Einstein's contraction is kinematical rather than dynamical in nature as it is for Lorentz). How can these two things be reconciled? 

Einstein answered this question “with a comparison”. He said that  "It has to do with the second law of thermodynamics”, and Einstein related the impossibility of a Maxwell's Demon (perpetual motion of the second kind) to relativity and the contraction of lengths: “If one takes the assumption of the impossibility of a perpetuum mobile of the second kind as the starting point of the argument, then our law appears as almost an immediate consequence of the basic premise of the theory. But if one bases the theory of heat on the equations of motion of molecules, then our law appears as the result of a long series of most subtle arguments. […] So the above-mentioned points of view of Minkowski on the one hand, and of H. A. Lorentz, on the other, also seem to me completely justified” (\cite{CPAE3}). 

In an undated letter to Maurice Solovine Einstein writes: "The
method of the theory of relativity is analogous to the method of
thermodynamics; for the latter is nothing more than the systematic answer
to the question: how must the laws of nature be constructed in order to rule out the possibility of bringing about perpetual motion?" (Einstein to Solovine, undated letter, \cite{Einstein4}). And in 1919, Einstein had stressed that "the science of thermodynamics seeks by analytical means to deduce necessary conditions, which separate events have to satisfy, from the universally experienced fact that perpetual motion is impossible" (\cite{Einstein5}).  
Elsewhere, Einstein again maintains: “The most satisfactory situation is evidently to be found in cases where the new fundamental hypotheses are suggested by the world of experience itself. The hypothesis of the non-existence of perpetual motion as a basis for thermodynamics affords such an example of a fundamental hypothesis suggested by experience; […] In the same category, moreover, we find the fundamental hypotheses of the theory of relativity […]” (\cite{Einstein3}). 

Einstein has reduced the second law of thermodynamics to impossibility of a perpetuum mobile of the second kind (a Maxwell's demon). 
In his \emph{Autobiographical Notes}, Einstein speaks of the impossibility of a perpetuum mobile of the first and the second kind in the context of the discovery of the principle of relativity and the famous chasing a light beam thought experiment. In 1911, he wrote about an impossibility of a perpetuum mobile of the second kind (a Maxwell's demon) in the context of Lorentz's dynamical contraction and Minkowski's interpretation of his special theory of relativity. 

What can we learn from the above quotations? In my interpretation of the above quotations, according to Einstein, a perpetual motion machine of the first and the second type cannot exist, and in the same category, no signal transmission faster than light is possible. 
We can summarize this by saying that Einstein was occupied with \emph{impossibility statements} when speaking about principle theories, thermodynamics and relativity. I use here the phraseology of Aaronson: "Many of the deepest principles in physics are impossibility statements: for example, no superluminal signalling and no perpetual motion machines" (\cite{Aaronson}). 

Let us now examine the role of \emph{impossibility statements} in Maxwell's demon thought experiment. 
Maxwell’s efficient demon must have a very quick reaction to the passage of molecules (\cite{Finfgeld}). 
Maxwell writes to Strutt: “Of course he [the so-called demon] must be quick for the molecules are continually changing both their courses and their velocities” (\cite{Harman}). How quick must the demon be? To “allow only the swifter molecules to pass from A to B, and only the slower ones to pass from B to A”, the demon needs to acquire detailed information about as many as possible gas molecules at a certain time instant. The demon must accurately distinguish the molecules and separate the slow molecules from the fast ones; he must quickly count many molecules at the same time or “at the same instant”. That is, according to Maxwell, the demon can see individual molecules “at the same instant”. But it is impossible to measure the precise velocities of all molecules “at the same instant” because no signal transmission faster than light is possible. 

Again Maxwell wrote to Strutt that the demon is “exceedingly quick, with microscopic eyes but still an essentially finite being” (\cite{Harman}). Maxwell stresses that the “being whose faculties are so sharpened that he can follow every molecule in its course, such a being, whose attributes are still as essentially finite as our own” (\cite{Maxwell1}). That is, the being is not supernatural and is subject to the laws of physics. Because Maxwell’s demon is a “finite being”, and no signal transmission faster than light is possible, the demon cannot simultaneously count all molecules and complete the sorting of molecules in reasonable time; and a perpetual motion machine of the second kind cannot be realized.  

Finally, before the advent of special relativity, in 1900, Karl Pearson suggested in his book, \emph{The Grammar of Science}, a so-called relativistic version of the Maxwell demon: “IRREVERSIBILITY of natural processes is a purely relative conception. History goes forward or backward according to the relative motion of the events and their observer. Conceive a colleague of Clerk Maxwell's demon", says Pearson, "Now suppose him to travel away from our earth with a velocity greater than that of light. Clearly all natural processes and all history would for him be reversed. Men would enter life by death, would grow younger and leave it finally by birth". This rings a bell. I suppose you know the story, \emph{The Curious Case of Benjamin Button}, by Francis Scott Fitzgerald from 1922. Benjamin is born in 1860, a 70-year-old man with white hair and a long beard. Benjamin grows backward at a normal rate, enters kindergarten at 65, goes through school and keeps growing younger until he cannot walk or talk.     
Pearson adds that the "conception of historical change and of time as a problem in relative motion" was suggested to him by Loius Napoleon George Filon, "and is, I think, of much interest from the standpoint of the pure relativity of all phenomena” (\cite{Pearson}). 
Three years before publishing his special theory of relativity in 1905, Einstein read Pearson’s book.

\section*{Acknowledgement}

This work is supported by ERC advanced grant number 834735.

\end{document}